\documentclass[11pt,twoside]{article}

\usepackage{asp2006}
\usepackage{epsf}
\usepackage{psfig}
\usepackage{lscape}
\usepackage{graphicx}

\begin{document}
\title{Can Terrestrial Planets Form in Hot-Jupiter Systems?}
\author{Martyn J. Fogg and Richard P. Nelson.}
\affil{Astronomy Unit, Queen Mary, University of London, Mile End
Road, London, E1 4NS.}

\begin{abstract}
Models of terrestrial planet formation in the presence of a
migrating giant planet have challenged the notion that hot-Jupiter
systems lack terrestrial planets. We briefly review this issue and
suggest that hot-Jupiter systems should be prime targets for future
observational missions designed to detect Earth-sized and
potentially habitable worlds.
\end{abstract}



\section{Introduction.}

Since the discovery of the first extrasolar planets
\citep{wolszczan,mayor}, astronomical techniques and observational
baselines have advanced to the point where over 200 extrasolar
planetary systems have been identified \citep{butler}. Most detected
exoplanets are in the giant planet mass range and it is now clear
that our solar system is but one variant within a great diversity of
planetary system architectures. One of the most surprising
discoveries has been of a population of giant planets, the so-called
\emph{hot-Jupiters}, found orbiting in a region of extreme
insolation very close ($r < 0.1~\mathrm{AU}$) to their central stars
and well within the radius of the original nebular snowline ($r
\approx 3 - 5~\mathrm{AU}$) where giant planets are thought to form
\citep{pollack}. Hot-Jupiters are not uncommon: they amount to about
a quarter of exoplanet discoveries, and are thought to provide
evidence that protoplanets can migrate over large radial distances
via tidal interactions with the protoplanetary disk
\citep[e.g.][]{lin1,lin2,ward2,nelson1}. Since the disk gas is
observed to disperse within the first few Myr of the system's
existence \citep{haisch}, giant planets must form and migrate
through the inner system within this period, which is considerably
less than the $\sim$~10--100~Myr thought to be required to complete
terrestrial planet formation
\citep{chambers2,kleine,halliday,obrien}.

Test particle studies have shown that terrestrial planets external
to a hot-Jupiter would have stable orbits \citep{jones}, and
\citet{raymond1} have shown that they should be able to form, in the
presence of a giant planet already at $\sim 0.1$~AU, from any
available pre-planetary material with a period ratio roughly $>$~3.
However, until recently it has been a common assumption that
terrestrial planets could not have formed in hot-Jupiter systems due
to the disruptive effect of the giant planet's migration which is
deemed to have cleared the inner system of planet-forming material
\citep[e.g.][]{armitage1}, prompting claims that the observed
abundance of hot-Jupiters could be used to constrain the general
abundance of habitable planets \citep{ward1}, and even their
galactic location \citep{lineweaver1}.

This picture has been challenged by the work of two groups who have
modeled terrestrial planet formation concurrently with, and
following, an episode of giant planet migration
\citep{fogg1,fogg2,fogg3,fogg4,raymond2,mandell}. Their findings
suggest that inner system solids disks are \emph{not} destroyed by
the intrusion of a migrating giant planet and that terrestrial
planet formation can resume in the aftermath and run to completion.

In this paper, we briefly describe our model of terrestrial planet
formation and show some typical results.

\section{The Model}

We picture our model systems to be composed of an inner system
solids disk, at its oligarchic growth stage \citep{kokubo1},
extending between 0.4 -- 4.0~AU, with its initial parameters being
those of a classic minimum mass solar nebula (MMSN) model
\citep{hayashi}, scaled up in mass by a factor of 3. This disk is
embedded in the nebular gas which depletes over time by accreting
onto the central star. We generate a number of different scenarios
by allowing planetary growth to proceed in the inner disk for
between 0.1 -- 1.5 Myr before placing a 0.5~$\mathrm{M_{Jup}}$ giant
planet at 5.0~AU and allowing it to migrate inward. We end the
migration episode when the giant planet reaches 0.1~AU. Since our
giant is massive enough to open up a gap in the gas disk, we are
specifically assuming the operation of type II migration, where the
planet is locked into the viscous evolution of the gas and migrates
inward in step with its radial motion
\citep[e.g.][]{lin1,lin2,ward2,nelson1}.

We use a modified version of the \emph{Mercury 6} hybrid-symplectic
integrator to perform our calculations \citep{chambers1}, and run it
as an $N + N'$ simulation, where we start with $N$ protoplanets
embedded in a swarm of $N'$ ``super-planetesimals". These latter
objects are tracer particles with masses a tenth of the initial
masses of protoplanets that act as an idealized ensemble of a much
larger number of real planetesimals and are capable of exerting
realistic dynamical friction on larger bodies
\citep[e.g.][]{thommes1}. The central star, giant planet, and
protoplanets interact gravitationally and can accrete and merge
inelastically with all other bodies. Super-planetesimals however are
non-self-interacting but subject to gas drag equivalent to that
experienced by a single 10~km radius planetesimal. Details of these
aspects of our model are given in \citet{fogg1}.

We calculate the evolution of the nebular gas using a 1-D viscous
disk model, with an alpha viscosity of $\alpha = 2\times10^{-3}$,
that solves numerically a modified viscous gas disk diffusion
equation that includes the tidal torques exerted by an embedded
giant planet \citep{lin1,takeuchi} and have described its
implementation in \citet{fogg3}. The gas responds over time by
draining via viscous accretion onto the central star; opening up an
annular gap centered on the giant planet's orbit; and forming a
partial inner cavity due to dissipation of propagating spiral waves
excited by the giant planet. The back reaction of these effects on
the giant planet is resolved as torques which self-consistently
drive type II migration.

\section{An Example Scenario.}

\begin{figure}
\centering
 \includegraphics[width=14cm,draft=false,keepaspectratio=true]{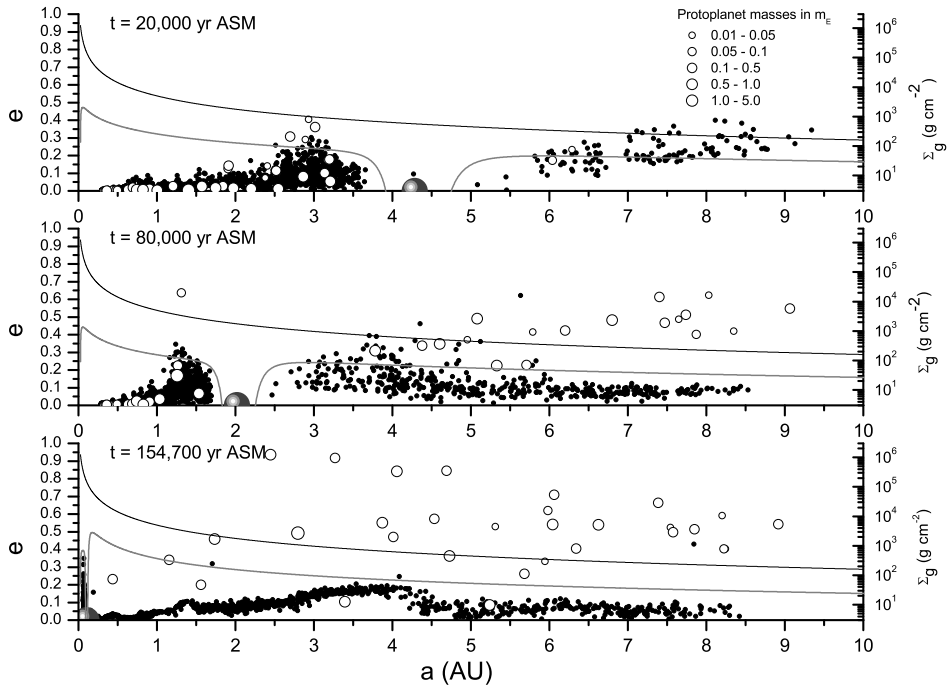}

\caption{Details of a migration scenario through a 1~Myr old inner
disk. The three panels, from the top down, show the system at
20\,000, 80\,000 and 154\,700 years after the start of giant planet
migration (ASM). Eccentricity of objects is plotted vs. semi-major
axis and is read from the left hand axes. Black dots are
super-planetesimals, grey or white circles are protoplanets, and the
large highlighted circle is the giant planet. Gas surface density is
read from the right hand axes, with the upper black curve being the
initial profile for a 3$\times$MMSN and the lower grey curve being
the current evolved profile.}
 \label{figure:1}
\end{figure}

We illustrate the behavior of one of our scenarios in
Fig.~\ref{figure:1}, where the system has been previously evolved
for 1~Myr before the appearance of the giant planet and the start of
its migration. Three time slices from within the migration episode
are shown in the three panels: the lowest panel showing the point at
which the giant planet halts at 0.1~AU, 154\,000 years after the
start of migration.

The results show that the passage of the giant does not sweep the
inner system clear of planet-forming material. Instead, the giant
planet shepherds the solids disk inward, compacting it and exciting
the orbits of objects captured at mean-motion resonances. Much of
this excited material eventually experiences a close encounter with
the giant planet and is expelled into an exterior orbit, augmenting
a new disk of solid material that progressively builds up in orbits
external to the final position of the hot-Jupiter. In this
particular case, 86\% of the solids disk survives, with 82\% of it
residing in the external scattered disk. Mass loss is modest and
mostly occurs via accretion onto the giant planet. The effect of the
giant's passage is therefore not the elimination of the inner system
disk, but instead a modest dilution and strong excitation of solid
material and a radial mixing that drives volatile-rich material
inward. This outcome is robust to considerable variation of model
parameters, such as the mass of the giant and the timing of its
formation and migration. The results of further simulation of
accretion in this scattered disk show that the initially eccentric
orbits of protoplanets are rapidly damped and circularized via
dynamical friction exerted by smaller bodies and possibly via tidal
drag exerted by the remaining gas \citep{fogg4}. Planetary growth
resumes and over the following $\sim$~10 -- 100~Myr gives rise to a
set of water-rich terrestrial planets in stable orbits external to
the hot-Jupiter \citep[see also][]{raymond2,mandell}.

\section{Conclusions.}

Our models predict that terrestrial planets might be routinely
expected in hot-Jupiter systems, including within their habitable
zones, and may be detectable by forthcoming missions such as Kepler,
Darwin, SIM PlanetQuest and TPF.

\end{document}